\documentclass[
]{ceurart}

\begin{document}

\copyrightyear{2021}
\copyrightclause{Copyright for this paper by its authors.
	Use permitted under Creative Commons License Attribution 4.0
	International (CC BY 4.0).}

\conference{CLEF 2021 -- Conference and Labs of the Evaluation Forum, 
	September 21--24, 2021, Bucharest, Romania}

%%
%% The
%\conference{CLEF'21: Conference and Labs of the Evaluation Forum,
%	September 21--24, 2021, Bucharest, Romania}

%% Do not remove the subtitle Notebook for the Touch{\'e} Lab on Argument Retrieval at CLEF 2021
\title{\texorpdfstring{\centering}{\texttwoinferior} Retrieving Comparative Arguments using \\ Ensemble Methods and Neural Information Retrieval}
\title[mode=sub]{\texorpdfstring{\hskip}{\texttwoinferior} 4em Notebook for the Touche Lab on Argument Retrieval at CLEF 2021}

\author[1,2]{Viktoriia Chekalina}

\author[1]{Alexander Panchenko}
\address[1]{Skolkovo Institute of Science and Technology, Moscow, Russian Federation}
\address[2]{Philips Innovation Lab Rus, Moscow, Russian Federation}

\begin{abstract}
  In this paper, we present a submission to the Touché lab's Task 2 on Argument Retrieval for Comparative Questions~\cite{bondarenko:2021a, bondarenko:2021c}. %For the set of given topics, we retrieve
%the corresponding responses and rank them using proposed approaches.
Our team Katana supplies several approaches based on decision tree ensembles algorithms to rank comparative documents in accordance with their relevance and argumentative support. We use PyTerrier~\cite{ecir2021-tut-bow2b} library to apply ensembles models to a ranking problem, considering statistical text features and features based on comparative structures. We also employ large contextualized language modelling techniques, such as BERT~\cite{devlin2019bert}, to solve the proposed ranking task. To merge this technique with ranking modelling, we leverage neural ranking library OpenNIR~\cite{macavaney:wsdm2020-onir}. 

Our systems substantially outperforming the proposed baseline and scored first in relevance and second in quality according to the official metrics of the competition (for measure NDCG@5 score). Presented models could help to improve the performance of processing comparative queries in information retrieval and dialogue systems.

%The quality metric result is only marginally different from Top 1.%

%We present a system scored second in the competition substantially outperforming the proposed baseline according to both official metrics. 

\end{abstract}


\maketitle

\section{Introduction}

On a daily basis, people face the problem of choosing between two entities - which phone is more reliable, which juice contains less sugar, which hotel is better for a holiday. Domain-specific comparison systems, like \href{https://www.wolframalpha.com/}{WolframAlpha} or \href{https://www.diffen.com/}{Diffen}, solve this problem partly and rely on structured data, which limits the number of cases it can be used. 

On the other hand, the Web contains a vast number of opinions and objective arguments that can facilitate the comparative decision-making process. It creates the need of developing an open-domain general system that could process such information. The issue is to retrieve from a set of documents relevant, supportive and credible arguments. The aim of the proposed work is to retrieve from ClueWeb12~\footnote{\url{http://lemurproject.org/clueweb12}} corpus documents and re-rank them, considering argumentation for or against one option or the other.

The contribution of our work is the following: we are first to use ensemble methods based on mixed statistical and comparative features to the document ranking; we are first to use neural information retrieval approach to the task of argument retrieval; we propose a model outperforming the baseline and yielding the first and the second-best result according to the relevance and quality metric, respectively.

\section{Related work}

The most relevant to this work is the previous shared task Touche 2020~\cite{touche20}. 17 participants took part in the competition and submitted 41 runs. Various approaches were tested by these participants, including methods based on extraction of structures corresponding to claims and premises, assessing argument quality, representation of documents by language models, expansion of the query by similar words. The ranking function from search engine ChatNoir~\cite{ChatNoir} based on BM25F~\cite{25bmf} approach was used as a baseline.

Only a few of the submitted solutions can slightly improve the baseline. The best overall approach in the previous competition was the method based on query extension and reranking documents by relevance, credibility, and supportive quality~\cite{Abye2020AnOW}.

This work is based on our run submitted in the previous version of the Touche shared task~\cite{Chekalina2020RetrievingCA}. In this work, we used a pre-trained language model to find relevance between the query and document. Extraction of comparative structures and counting the number of comparative sentences in a document help us to assess the quality of relevant arguments.

Therefore, the problem of argument retrieval arises in other scenarios. Comparative argumentation machine CAM~\cite{CAM} retrieves comparative sentences with respect to accepted objects and comparison aspects. The paper~\cite{Fromm} explores the influence of context on an argument detecting system and proves the performance increasing related to it.

\section{Datasets and experimental design}

\subsection{Datasets}

The organizers provided 50 comparative questions~(topics), for which we should obtain documents containing convincing arguments for or against one or another option. Topics for the competition are available online. \footnote{\url{https://webis.de/events/touche-21/shared-task-2.html}}

In addition, 50 topics and corresponding relevance annotations of the previous year's competition~\cite{10.1007/978-3-030-58219-7_26} were given for supervised learning. These documents were also retrieved from ChatNoir and ranked manually to 0~(not relevant), 1~(relevant) or 2~(highly relevant) scores. We use this data to train and set up models based on the decision trees and fine-tune the BERT ranker. Besides, last year's teams submissions were available too.

Unfortunately, this data is not insufficient for fitting large supervised ranking models, for example, based on the BERT technique. In this case we use adjacent question-answering dataset called Antique~\cite{DBLP:journals/corr/abs-1905-08957}. This dataset consists of the questions and answers of Yahoo! Webscope L6 and contains 2,626 open-domain non-factoid questions and 34,011 manual relevance annotations. 

The example of query and ranked answers are in Table~\ref{tab:touche_example}, Table~\ref{tab:antique_example} in Appendix A. It might be noticed that Antique dataset has a different set of ranking scores - {0,~1,~2} instead of {1,~2,~3,~4} - so we rewrite Antique ranks in accordance with the following mapping 1$\rightarrow$0, 2$\rightarrow$1, 3$\rightarrow$1, 4$\rightarrow$2.

\subsection{Evaluation setup}

We use every topic as a query in ChatNoir~\cite{ChatNoir} search engine and extract up to 100 unique documents from the ClueWeb12 corpus. We clean documents’ bodies from HTML tags and markups and ranked them using one of the developed approaches described below.

As auxiliary data, the organizers provided the topics of the previous year's competition. For each proposed topic, a set of documents from ChatNoir was retrieved and labelled as described above.
We use this data to train developing models and valid composing approaches. In the validation phase, we split the ranked data into 40 topics in train and 10 in validation. 

In the run phase, we execute produced solutions on web evaluation platform Tira~\cite{TIRA}. In this stage to fit the model we use ranked data from the previous year entirely and predict rank for current proposed topics.
The runs were evaluated using the NDCG metrics based on the human judgements of the submitted runs. Retrieved documents were judged in accordance with two criteria: (i) document relevance, (ii) whether sufficient argumentative support is provided~\cite{supporting_human_for}.

\section{Document ranking using ensembles of trees}

In this section, we use ensembles of trees as a supervised machine learning technique to solve ranking problems. We choose either pointwise regression tree algorithms, like Random Forest, or boosted tree algorithms like XGBoost and LightGBM. 
In the cases of LightGBM model we employ LambdaMART~\cite{wu2010adapting} objective. It combines cost function derived from minimizing the number of inversions in ranking~(LambdaRank~\cite{burges2007learning}) and objective for building gradient boosted decision trees~(MART~\cite{article}).
We use PyTerrier platform for information retrieval.\footnote{\url{https://pyterrier.readthedocs.io/en/latest/index.html}} It simplifies the extraction of the text features and allows expressing retrieval experiments~\cite{pyterrier2020ictir}.

\subsection{Feature extraction}

For our ranking ML methods, we use features that came from 3 origins described below:  (i) ranking features extracted by PyTerrier, (ii) specific comparative features, (iii) score from ChatNoir system based
on custom BM25 scoring function.\footnote{\url{https://www.elastic.co/guide/en/elasticsearch/reference/current/index-modules-similarity.html}} 

\subsubsection{Features extracted by PyTerrier}

PyTerrier provides measure of matching query-document texts by several models. Among these models are statistical measures~(TF-IDF), mesures based on language models~(Heimstra, Diriclet), measures based on occurrence of a document depending on the fields that the term occurs in~(BM25F, PLF).
The list of all possible models are available at the cite~\footnote{\url{http://terrier.org/docs/current/javadoc/org/terrier/matching/models/package-summary.html}}. Among these varieties we have chosen BM25, Heimstra, DFIC, DPH, TF-IDF, DiricletLM, PL2 for our exploration. 

\newpage

\newlength{\oldintextsep}
\setlength{\oldintextsep}{\intextsep}

\setlength\intextsep{0pt}
\begin{wraptable}{l}{4.5cm}
\hspace{0.15mm}
\fontsize{10pt}{10.5pt}\selectfont
\caption{Results on validation set for text features in PyTerrier models.}
\centering
\label{preprocessing1}
\begin{tabular}{| l | c |}
\toprule
Method -Score & NDCG@5  \\
\midrule
\textbf{BM25} & \textbf{0.3637} \\ 
Heimstra & 0.3616 \\
\textbf{DFIC} & \textbf{0.3642}\\
DPH & 0.3110 \\
\textbf{TF-IDF} & \textbf{0.3637} \\
DiricletLM & 0.3307 \\
PL2 & 0.3703 \\
\bottomrule
\end{tabular}
\end{wraptable}

We applied each of the selected methods sequentially and independently to the training set, ranked documents by the obtained scores and evaluated the ranking on the validation set. The result of these tests is in Table~\ref{preprocessing1}. We have chosen 3 methods with the most promising results, and these 3 methods combine 3 features.

\subsubsection{Comparative features}

We focus not only on finding high relevant documents as on finding documents with a comparison of one object relative to another.
The work~\cite{chekalina-etal-2021-better} assumes that the comparative issue can be represented by comparative structures - objects for comparison, comparative aspects and predicates. We take the sequence-labelling model suggested in the cited paper and applied it to the query. It helps us to define objects for comparison for every topic. Then we apply the model to document and get a comparative feature set.

The feature {\tt is\textunderscore retrieved} describes are there any comparative structures in the document at all. Characteristic {\tt objs\textunderscore score} defines how many objects from query are found in document (0, 1 or 2). Feature {\tt asp\textunderscore pred\textunderscore score} is counted in the following way: if at least one object from a query is in the document, every word in the document labelled as an aspect or predicate increases the score to 0.5. Finally, we combined defined features with scores obtained from the ChatNoir system, and a resulting feature vector for pair query-document is \{{\tt score\textunderscore pl2}, {\tt score\textunderscore tf}, {\tt score\textunderscore bm}, {\tt score\textunderscore dfic}, {\tt baseline\textunderscore scores}, {\tt is\textunderscore retrieved}, {\tt ap\textunderscore score}, {\tt objs\textunderscore score}\}.
\subsection{Models}
\subsubsection{Random Forest}

We use the Random Forest model imported from Sklearn and wrapped by the PyTerrier pipeline. To find the best setup, we vary the number of estimators from 10 to 150, the value 20 gives the best valid score NDCG@5 0.408.

\subsubsection{XGBoost}

We also wrapped gradient boosting library from Sklearn to PyTerrier class and tune hyperparameters by setting the learning rate from $1{\rm e}{-4}$ to 0.1 and max\textunderscore depth from 4 to 16. 
The best setup is learning rate $0.01$, max\textunderscore depth $6$ and gives NDCG@5 0.547.

\subsubsection{LightGBM}

%\begin{wraptable}{l}{4.5cm}
%\caption{Feature importance in the proposed LightGBM model}
%\centering
%\label{table:features_lgbm}
%\begin{tabular}{| l | c |}
%\toprule
%Feature & Importance  \\
%\midrule
%PL2 & 1.76 \\ 
%Heimstra & 0.3616 \\
%TF-IDF & 1.19\\
%DFIC & 2.3 \\
%Chat Noir & 20.8 \\
%has comp & 0 \\
%objs\_score & 1.66 \\
%asp\_pred & 1.51 \\
%\bottomrule
%\end{tabular}
%\end{wraptable}

In the case of LightGBM, we vary the number of leaves from 8 to 20 and the learning rate from 0.001 to 0.1. The best configuration with num\_leaves = 15 and learning rate = 0.1 gives 0.579 score.

The feature importance of the resulting model is in Table~\ref{table:features_lgbm}. It can be seen that the most significant feature is the score retrieved from the ChatNoir, then there is a Divergence from Independence based on Chi-square~\cite{DFIC} and the existence of comparison objects in the document.

\section{Document ranking using neural information retrieval based on BERT}

Contextualized language models such as BERT can be much more efficient for ranking tasks because they contain vast relationships between language units. In the proposed work we use a reranking model from OpenNIR~\cite{macavaney:wsdm2020-onir} based on ``Vanilla'' Transformer architecture.

\subsection{Text representation}
BERT receives a query and document and processes it jointly. A distinctive feature of the BERT reranker is injection token similarity matrices on each layer, which considerably improves performance~\cite{MacAvaney_2019}.

\begin{table}[ht]
\caption{Feature importance in the proposed LightGBM model} % title of Table
\centering % used for centering table
\begin{tabular}{|c | c c c c c c c c|} % centered columns (4 columns)
\toprule %inserts double horizontal lines
Feature & Pl2 & TF-IDF & BM25 & Dfic & ChatNoir & has comp & objs\_score & asp\_pred \\ [0.5ex] % inserts table
\midrule % inserts single horizontal line
Importance & 1.76 & 1.19 & 1.51 & 2.3 & 20.8 & 0 & 1.66 & 1.51 \\ \bottomrule % inserting body of the table

\end{tabular}
\label{table:features_lgbm} % is used to refer this table in the text
\end{table}

\subsection{Training process}

First, we pretrain this reranker on the Antique dataset. We clean this dataset from incorrect symbols and makeups. We also left from the dataset documents of length more than 300 characters, since the length of the ChatNoir retrieves usually does not exceed 300. The training process lasted for 500 epochs with 0.001 learning rate and 56 objects in every batch. Finally, our model gives NDCG@5 0.3362 on a validation set. We fine-tune the model on 40 train topics from the previous year for 50 epochs with the same configuration. Fine-tuning increased the score on validation up to 0.412.

%\begin{minipages}
%\begin{minipage}{0.5\textwidth}

%We define comparative structures: objects, predicates, aspects. To fit sequence-labelling models, we create Comparely, a manually labeled dataset with comparative sentences.

%\end{minipage}
%\hspace{0.15mm}
%\vspace{-0.25mm}
%\begin{minipage}{0.43\textwidth}
%   \begin{table}
%\caption{Ranking scores on valid set for text features in PyTerrier packages}
%\centering
%\label{preprocessing1}
%\begin{tabular}{| l | c |}
%\toprule
%Method -Score & NDCG@5  \\
%\midrule
%\textbf{BM25} & \textbf{0.3637} \\ 
%Heimstra & 0.3616 \\
%\textbf{DFIC} & \textbf{0.3642}\\
%DPH & 0.3110 \\
%\textbf{TF-IDF} & \textbf{0.3637} \\
%DiricletLM & 0.3307 \\
%PL2 & 0.3703 \\
%\bottomrule
%\end{tabular}
%\end{table}
%\end{minipage}

%\end{minipages}

\section{Results}

\subsection{Results on validation set}
%\begin{wraptable}{l}{6.5cm}

%\newlength{\oldintextsep}
%\setlength{\oldintextsep}{\intextsep}

\setlength\intextsep{0pt}
\begin{wraptable}{l}{5.8cm}
\fontsize{9pt}{9.5pt}\selectfont
\caption{Results on validation set.}
\centering
\label{valid1}
\begin{tabular}{| l | c | c |}
\toprule
Method & NDCG@5 & Time, ms  \\
\midrule
Random Forest & 0.408 & 127.168  \\ 
XGBoost& 0.547 & 128.848\\
\textbf{LightGBM} & \textbf{0.572} & 131.244\\
\midrule
Bert Ranker & 0.412 & 1560.947\\
\midrule
Baseline'20 & 0.534 & -\\
%B.Beggins (Top 1) & 0.517 & -\\
%Top 1 (B.Beggins) & 0.400 & -\\
\bottomrule
\end{tabular}
\end{wraptable}

% !!! in Time Bert without pre-trained on Atique, only fine-tune

%\end{wraptable}
The result for every proposed approach obtained on the validation part of data from the previous year competition is in Table~\ref{valid1}. We also evaluate the previous year's baseline on the validation set. % and Top 1 methods~\cite{10.1007/978-3-030-58219-7_26}. % and winner team Bilbo Baggins.
The best scores come from the LightGBM model, which also outperforms the baseline. % and Top 1.
XGBoost has fewer scores, Random Forest as a simple algorithm has the smallest score. Bert overtakes Random Forest a little. 

In the right column, we also added the time required to train each model. It can be seen that the ensemble-based models have approximately the same time complexity, while the Bert requires much more time to train.

%\newlength{\oldintextsep}
%\setlength{\oldintextsep}{\intextsep}

%\setlength\intextsep{0pt}

\subsection{Results on test set}

For final testing, the retrieved documents were labelled manually with a score from 0 to 3. Judgment was carried out in two independent criteria: the relevance of the document to the given topic and the quality of the text. Quality criterion includes good language styling, easy reading and proper sentence structure, the absence of typos and alliteration.

For each criterion, a separate file with the assessor's scores is available.
The results of two evaluations are presented in the Table~\ref{runs_relevance} and Table~\ref{runs_quality}. The runs of our team Katana have the best result between all teams in terms of relevance and the second result in terms of the text quality. 

As in the validation set, XGBoost and LightGBM give the best performance. It is well explained, since the loss of these models based on the ranking quality functions, NDCG in the XGBoost case and LambdaMART in the LightGBM case. The first model describes relevance a bit better (0.489) and has first place among the whole participant. For quality, conversely, LightGBM is better. It archives 0.684 and takes second place in a quality table, slightly surrendering to Top 1. The random forest method has scores just below the baseline in both cases. It can be explained by a more elementary algorithm for building an ensemble. Bert gives a quite good result for quality and weak for relevance. Perhaps the data from the adjacent task~(factoid QA) used for the training is the reason for not a very accurate solution. %Perhaps the reason was the insufficiency of data for learning for such an extensive model.

\begin{table}[!htb]
    %\caption{NDCG@5 scores on test set for approaches}
    %\label{runs1}
    \begin{minipage}{.5\linewidth}
      \centering
        \caption{NDCG@5 scores on runs for relevance for Katana \\ team, baseline and Top-2 approach
        }
        \label{runs_relevance}
        \begin{tabular}{| l | c |}
        \toprule
        Method & NDCG@5  \\
        \midrule
        Random Forest & 0.393 \\ 
        \textbf{XGBoost~(Top 1)}& \textbf{0.489} \\
        LightGBM & 0.460 \\
        \midrule
        Bert Ranker & 0.091 \\
        \midrule
        ChatNoir baseline & 0.422 \\
        Thor team~(Top 2) & 0.478 \\
        \bottomrule
        \end{tabular}
    \end{minipage}%
    \begin{minipage}{.5\linewidth}
      \centering
        \caption{NDCG@5 scores on runs for quality for Katana \\ team, baseline and Top-1 approach}
        \label{runs_quality}
        \begin{tabular}{| l | c |}
        \toprule
        Method & NDCG@5  \\
        \midrule
        Random Forest & 0.630\\ 
        XGBoost& 0.675 \\
        \textbf{LightGBM~(Top 2)} & \textbf{0.684}\\
        \midrule
        Bert Ranker & 0.466\\
        \midrule
        ChatNoir baseline & 0.636\\
        Rayla team~(Top 1) & 0.688\\

        \bottomrule
        \end{tabular}
    \end{minipage}%
\end{table}

%Under the terms of the experiment, evaluation of the final results will be after the time of the paper submission. We set Table~\ref{runs1} for final results and will fill void cells after the evaluation.

% the relevance of results, but also the
%argumentative quality.

\section{Conclusion}
In this paper, we present our solution to the Argument retrieval shared task. We pay attention to ensembles methods and use statistic approaches, language modelling and comparative structure extraction to retrieve features for it. We also use a neural reranker based on the Bert technique to use information from a contextualized model in our task. 

The best results were obtained by gradient boosting methods, training on ranking cost function: XGBoost and LightGBM. The proposed approaches outperform baseline and take first and second places in relevance and quality ranking, respectively. Bert contextualized model shows the need for large learning data.

\clearpage

\begin{acknowledgments}
This work is partially supported by the project ``ACQuA:\,Answering Comparative Questions with Arguments'' (grants BI~1544/7-1 and HA~5851/2-1) as part of the priority program ``RATIO: Robust Argumentation Machines'' (SPP 1999). We thank Maik Frobe for providing the support of the software runs in the TIRA system.

\end{acknowledgments}

\section*{Appendix A: Examples of training data}

\begin{table}[ht]
    \centering
    \caption{Example of query and document with different relevance in Touche task dataset}
    \begin{tabular}{|p{0.20\linewidth} | p{0.7\linewidth} | p{0.06\linewidth}|} 
    \toprule
      Query & Document & Rank \\ \midrule
      \multirow{3}{*}{\shortstack[l]{What is better\\ for the environment,\\a real or a fake\\Christmas tree?}}
       & Disease and condition content is reviewed by our medical review board real or artificial? There is so much confusing information out there about which is better for your health and the environment. & 2\\ 
       \cline{2-3}
        & You may think you're saving a tree, but the plastic alternative has problems too. Which is ``greener'' an artificial Christmas tree or a real one? & 1 \\ 
        \cline{2-3}
       & This entry is part 25 of 103 in the series eco-friendly friday november 28th’s tip christmas trees: stuck between choosing a real Christmas tree or a fake one? & 0 \\ \bottomrule
    \end{tabular}
    \label{tab:touche_example}
\end{table}

\begin{table}[ht]
    \centering
    \caption{Example of query and document with different relevance in Antique dataset}
    \begin{tabular}{|p{0.20\linewidth} | p{0.7\linewidth} | p{0.06\linewidth}|}
    \toprule
      Query & Document & Rank \\ \midrule
      \multirow{3}{*}{\shortstack[l]{Why do we put the \\ letter k on the words \\ knife and knob, knee?}}
      & They are saxon words. Knife would have been pronounced ker-niff. & 4\\ \cline{2-3}
       & As a guess I would say that historically ``kn'' would have been pronounced differently to ``n'' and that time has altered the way the words are pronounced. & 3 \\ \cline{2-3}
       & Because English is a funny language. & 2 \\ \cline{2-3}
       & I don't really (k)now! & 1 \\ \bottomrule
    \end{tabular}
    \label{tab:antique_example}
\end{table}

\section*{Appendix B: Examples of ranking results}

In this appendix you can find examples Top-3 ranked documents in accordance to LightGBM and Baseline approaches.

\begin{table}[!ht]
    \centering
    \caption{Example of documents with the different relevance to query ``Is admission rate in Stanford higher than that of MIT~?''}
    \begin{tabular}{| p{0.5\linewidth} | p{0.5\linewidth}|}
    \toprule
    \multicolumn{2}{|c|}{Is admission rate in Stanford higher than that of MIT~?} \\
    \midrule
      LightGBM Top-3 & Baseline Top-3 \\ \midrule
        1. Stanford and Harvard have a similar admissions rate of about 7\%. MIT comes with a somewhat greater rate of success admitting just under 10\% or 1742 for the class of 2015. Harvard, Stanford and MIT are global leaders in culture, commerce and governmental policies. & 1. Stanford and Harvard have a similar admissions rate of about 7\%. MIT comes with a somewhat greater rate of success admitting just under 10\% or 1742 for the class of 2015. Harvard, Stanford and MIT are global leaders in culture, commerce and governmental policies\\
      \midrule
       2. For more than a decade, i have served as an admissions officer for MIT. In that time, i've read more than 10,000 applications and have watched thousands of new students enter MIT. It is a privilege to work at the most dynamic and exciting university in the world. & 2. For more than a decade, i have served as an admissions officer for MIT. In that time, i've read more than 10,000 applications and have watched thousands of new students enter MIT. It is a privilege to work at the most dynamic and exciting university in the world. \\
      \midrule
      3. Our primary enhancement was targeted at families earning less than \$75,000 — making mit tuition free and eliminating  & 3. All of this factual information, plus a lot of other detail, can be found in the mit admissions literature. In fact, this year, mit will award \$74 million in undergraduate aid. \\
      \bottomrule
    \end{tabular}
    \label{tab:stanford_mit}
\end{table}

\begin{table}[!ht]
    \centering
    \caption{Example documents with the different relevance to query ``Which smartphone has a better battery life: Xperia or iPhone~?''}
    \begin{tabular}{| p{0.5\linewidth} | p{0.5\linewidth}|}
    \toprule
    \multicolumn{2}{|c|}{Which smartphone has a better battery life: Xperia or iPhone~?} \\
    \midrule
      LightGBM Top-3 & Baseline Top-3 \\ \midrule
        1. 1. The power saver app that will turn down settings when battery life is low to get as much juice out of the battery as possible. Sony has set the benchmark with its 12 megapixel camera inside the Xperia S. & 1. The iPhone 4 is apple's thinnest smartphone yet, but offers a much better screen, faster processor, video calling, and many other enhancements.\\
      \midrule
       2. How to increase the battery life of apple's iPhone 4s many of those with an iphone 4s have complaints about the battery life. Apple has acknowledged these problems, and is working to fix them. & 2. Sony Xperia's review: an above average smartphone ~ gizmotraker', as far as battery life is concerned, it last about 7 hr 30 min in talktime, 450 hrs in standby. \\
      \midrule
      3. Sony Ericsson includes an 8gb card in the sales package the Sony Ericsson Xperia arc s has below average battery life. Most users will get around 24 hours of life out of the Xperia. X27's 1600mah battery before it needs a recharge, but heavy users may need an injection of power before then. & 3. How to increase the battery life of Apple's Iphone 4s many of those with an iphone 4s have complaints about the battery life. Apple has acknowledged these problems, and is working to fix them. \\
      \bottomrule
    \end{tabular}
    \label{tab:battery}
\end{table}


%%
%% Define the bibliography file to be used

\clearpage
\bibliography{touche}

\appendix

\end{document}